\documentclass[12pt]{amsart}
\usepackage{amsbsy,amssymb,amscd,amsfonts,latexsym,amstext,delarray,
amsmath,graphicx} 
\input xypic

\usepackage{color}

\newtheorem{thm}{Theorem}[section]
\newtheorem{prop}[thm]{Proposition}
\newtheorem{cor}[thm]{Corollary}
\newtheorem{lem}[thm]{Lemma}
\newtheorem{conj}{Conjecture}[section]

\newtheorem{defn}[thm]{Definition}
\newtheorem{rem}[thm]{Remark}

\newtheorem{ques}[thm]{Question}
\numberwithin{equation}{section}

\def\bP{{\mathbb P}}

\def\F{{\mathbb F}}

\def\N{{\mathbb N}}

\def\Q{{\mathbb Q}}
\def\R{{\mathbb R}}
\def\Z{{\mathbb Z}}

\def\cA{{\mathcal A}}

\def\cC{{\mathcal C}}

\def\cO{{\mathcal O}}
\def\cP{{\mathcal P}}

\def\cR{{\mathcal R}}
\def\cS{{\mathcal S}}

\def\cU{{\mathcal U}}

\title{Computability questions in the sphere packing problem}
\author{Yuri I.~Manin and Matilde Marcolli}
\date{2022}
\address{Max Planck Institute for Mathematics, Bonn, D-53111, Germany}
\email{manin@mpim-bonn.mpg.de}
\address{Department of Mathematics and Department of Computing and Mathematical Sciences, 
California Institute of Technology, Pasadena, CA 91125, USA}
\email{matilde@caltech.edu}

\begin{document}
\maketitle
\begin{abstract}
We consider the sets of dimensions for which there is an optimal sphere packing
with special regularity properties (respectively, a lattice, or a periodic set with a 
given bound on the number of translations, or an arbitrary periodic set). We show
that all these sets are oracle-computable, given an oracle that orders 
an associated set of spherical codes by increasing Kolmogorov complexity. 
\end{abstract}

\section{Introduction}

A sphere packing $\cP\subset \R^n$ is an arrangement of non-overlapping spheres 
$S^{n-1}$ of equal sizes. The sphere packing problem is the question of identifying 
the packing, in a given dimension $n$, that achieves the maximal possible density,
that is, that maximizes the fraction of volume in $\R^n$ that is occupied by spheres
of the packing. We write $\Delta_\cP$ for the density of a given packing, while the
center-density $\delta_\cP$ is given by the density normalized by the volume of the unit sphere,
$$ \delta_\cP:=\Delta_\cP/{\rm Vol}(B_1^n(0)) \, . $$

\smallskip

A special case of sphere packings consists of {\em lattice packings}. These are sphere
packings for which there is a lattice $L\subset \R^n$ such that the spheres of the packing $\cP_L$
are centered at the lattice points. The sphere diameters are, in this case, equal to
the length $\ell_L$ of the shortest lattice vector. In the case of lattice sphere packings,
the density is given by
$$  \Delta_{\cP_L} = \frac{{\rm Vol}(B_1^n(0)}{| L |} \, \left(\frac{\ell_L^n}{2}\right)^n\, . $$
and the center-density by
\begin{equation}\label{centerdense}
 \delta_L =\left( \frac{\ell_L}{2} \right)^n \, \frac{1}{| L |} \, , 
\end{equation} 
with $\ell_L$ the shortest length of $L$. The covolume $| L |$ is assumed to be
fixed and can be taken equal to $1$, but we prefer to keep it explicit. 

\smallskip

It is known that, in general, the densest sphere packing need not be achieved by
a lattice packing. It is the case that the highest density is realized by a known lattice solution
in dimensions $1$, $2$, $3$, $8$, and $24$. These are the only explicitly known cases, with
the last three due, respectively, to \cite{Hales}, \cite{Viaz}, and \cite{CKMRV}. 

\smallskip

A larger special class of sphere packings is given by {\em periodic packings}. These are
sphere packings $\cP_\Sigma$ where the spheres are centered at the points of a 
{\em periodic set}, defined as follows.

\smallskip

\begin{defn}\label{perset}
A periodic set $\Sigma$ in $\R^n$ is a set for which there exist a finite collection
$\{ v_1, \ldots, v_N \}$ of vectors $v_i \in \R^n$ and a lattice $L \subset \R^n$
such that
$$ \Sigma = \cup_{i=1}^N v_i+ L \, . $$
The minimal number $N$ of translations describing the periodic set $\Sigma$
is the size $N=:\sigma(\Sigma)$ of $\Sigma$.
\end{defn}

\smallskip

Thus, a periodic set is a finite union of translates of a lattice
and a periodic sphere packing is a packing where a sphere is
placed at each point in a periodic set.

\smallskip

The density of a periodic packing with $N$ translations is given by
$$ \Delta_\Sigma= \frac{Vol_{\R^n}(B_1(0))\, N \, r^n}{|L|}\, , $$
with $|\Lambda|$ the lattice covolume and $r$ the sphere radius,
$r=\ell_\Sigma/2$, with 
$$ \ell_\Sigma= \min_{\lambda \in L, i,j=1,\ldots,N} \| \lambda +v_i - v_j \|\, . $$
The center density is correspondingly given by
\begin{equation}\label{centerdenseN}
 \delta_\Sigma =\frac{N \, \ell^n_\Sigma}{2^n |L|}\, . 
\end{equation}

\smallskip

It is expected that in very high dimension the densest sphere packings are given by
``disordered" rather than lattice solutions. However, the Zassenhaus conjecture 
expects that {\em in every dimension} $n$ the maximal density solution can be
realized by a {\em periodic packing}. 

\smallskip

In dimension $n=10$ it is known that the maximal density is realized by a periodic
packing involving $40$ translations, so this is the first dimension in which the known
solution is not a lattice solution. There are higher dimensional cases where the
maximal density is achieved by periodic non-lattice solutions. It is not known
whether there are dimensions in which the optimal density is realized by packings
that are not even periodic, though it is known
that periodic packings (where the number of translations is not bounded) can 
approximate with arbitrary precision any non-periodic packing. For dimension
$n\leq 8$ the maximal density is known and all the lattices that are local maxima
for density (though possibly not global solutions) are also known for  dimensions
$n\leq 8$, \cite{Schur}. All the local maxima of density given by periodic packing are
known for dimensions $n\leq 5$, for the case of periodic packings involving only
$2$ translations, \cite{AndKall}. These partial results show that it is in general
very difficult to gain information about how the densest solutions behave in
different dimensions. 

\medskip
\subsection{Decidability and the sphere packing problem}\label{CompIntro}

As mentioned above, there are very few dimensions for which the
sphere packing problem is solved. Moreover, there is no independent proof of
the decidability of the sphere packing problem itself, and the question
is open in any dimension in which the problem has currently not been solved.

\smallskip

Let $\cS\cP_n$ denote the set of all sphere packings $\cP\subset \R^n$
and let $\cS\cP=\cup_n \cS\cP_n$. Let  $\cS\cP^{\max}_n$ be the set of
sphere packings that realize the maximal possible density in $\R^n$. 
Note that, while in some
cases (such as dimension $2$, $8$ or $24$) there is a unique maximizer (after fixing the
lattice covolume), in other cases the maximizer is not unique: in dimension $3$ 
there are infinitely many distinct periodic packings of maximal 
density. Thus $\cS\cP_n^{\rm max}$ can in general be an infinite set, 
even when the scaling of the lattice is fixed. A related question is whether the
set $\cS\cP_n^{\rm max}$ is computable, namely whether there is an
algorithm that identifies and enumerates the elements of this set. 

\smallskip

In a similar way, we can consider the set  $\cS\cP^{{\rm Latt}}_n$
of lattice sphere packings $\cP_L\subset \R^n$ with $L\subset \R^n$ a
lattice, and $\cS\cP^{{\rm Latt}}=\cup_n \cS\cP^{{\rm Latt}}_n$. 
The {\em lattice-packing problem} is the question of finding, in a given 
dimension, a lattice that realizes the maximal density among lattice 
sphere packings. We write $\cS\cP^{{\rm Latt},\max}_n$ for the set
of lattice sphere packings $\cP_L\subset \R^n$ that achieve the
maximal density among lattice packings. 

\smallskip

Unlike the general sphere packing problem, the lattice-packing problem 
is a decidable problem, with an argument 
due to Voronoi showing that there are finitely many inequivalent local 
density maximizers and an algorithm that enumerates them. 
The sets $\cS\cP^{{\rm Latt},\max}_n$ are computable, and in fact
finite. 

\smallskip

A related problem, that we will be discussing in the following, 
is knowing for which dimensions the best lattice solution 
would also be the best solution among all sphere packings (as is known to be 
the case in dimensions $1,2,3,8,24$).

\smallskip

We can similarly consider the set  $\cS\cP^{{\rm Per}}_n$
of periodic sphere packings $\cP_\Sigma\subset \R^n$ with $\Sigma\subset \R^n$ a
periodic set, and $\cS\cP^{{\rm Per}}=\cup_n \cS\cP^{{\rm Per}}_n$, with
$\cS\cP^{{\rm Per},\max}_n$ the set of packings $\cP_\Sigma$ that achieve the
maximal density among all periodic packings. One can restrict the class of
periodic packings by considering only periodic sets with a fixed number of
translations of lattices. We write $\cS\cP^{{\rm Per},N}_n$ for the set of
periodic sphere packings of fixed size $\sigma(\Sigma)=N$ and 
$\cS\cP^{{\rm Per} \leq N}_n=\cup_{k=0}^N \cS\cP^{{\rm Per},k}_n$, with
$\cS\cP^{{\rm Per},0}_n=\cS\cP^{{\rm Latt}}_n$. Correspondingly, we write
$\cS\cP^{{\rm Per},N,\max}_n$ and $\cS\cP^{{\rm Per} \leq N,\max}_n$ for
the density-maximizer packings within these classes. 

\smallskip

It is known that, if one considers the class of sphere packings
given by periodic sets consisting of a fixed number $N$ of translations
of a lattice, then finding the density-optimizers in this class is a decidable
problem. The sets $\cS\cP^{{\rm Per},N,\max}_n$ and $\cS\cP^{{\rm Per} \leq N,\max}_n$
are computable (though they need not be finite in this case, as one knows
already in the case of dimension $3$). 

\smallskip

As in the lattice case, one can then ask what is the set of dimensions
for which the maximal density sphere packing is a periodic packing with
(up to) a fixed number $N$ of lattice translations.

\smallskip

The question of whether the packing problem is decidable in a
given dimension $n$, for arbitrary periodic packings in $\cS\cP^{{\rm Per}}_n$,
is also open for all dimensions where the packing problem is unsolved.
Correspondingly, one can ask whether the maximizer set 
$\cS\cP^{{\rm Per},\max}_n$ is computable. Correspondingly,
one also has a computability question about the set of dimensions
for which the maximal density sphere packing is a periodic packing
(with arbitrary size).

\smallskip

One can formulate additional questions of this kind. For
example, given a computable function $F:\N \to \N$, 
one can consider the set $\cS\cP^{{\rm Per}, F}_n$
of periodic packings $\cP_\Sigma\subset \R^n$ with
periodic set $\Sigma\subset \R^n$ of size $\sigma(\Sigma)=F(n)$,
and with $\cS\cP^{{\rm Per}, F}=\cup_n \cS\cP^{{\rm Per}, F}_n$.
In this case also the packing problem is decidable for packings
in this class. The maximizer sets $\cS\cP^{{\rm Per}, F,\max}_n$
are computable. So one can also investigate the set of
dimensions where the general sphere packing problem is
solved by a periodic packing in this class.  We formulate
these computability questions more precisely in the next subsection.

\medskip
\subsection{The sets of dimensions with lattice and periodic solutions}\label{OPTsec}

As above, we write $\cP_S$ for a sphere packing with the sphere centers located at
the points of the set $S\subset \R^n$. We also write $\Delta_{\cP}$ for the density
of a sphere packing $\cP$ and we write $\Delta_n^{\max}$ for the maximal sphere
packing density in dimension $n$,
\begin{equation}\label{Deltanmax}
\Delta_n^{\max}:= \sup_{\cP \subset \R^n}  \Delta_\cP\, ,
\end{equation}
with the supremum taken over all sphere packings $\cP$ in dimension $n$.
It is known that this maximal density is realized by some sphere packing, but
the nature of the packing that realizes the maximal density (lattice packing, periodic packing,
non-periodic disordered packing) is in general not known. 

\smallskip

\begin{defn}\label{nsetsSol}
We introduce the following sets of dimensions.
\begin{itemize}
\item The set ${\rm OPT}_{\rm Latt}$ is the set of dimensions $n\in \N$ such that the maximal
sphere packing density in dimension $n$ is realized by a lattice $L$,
\begin{equation}\label{OPTLatt}
{\rm OPT}_{\rm Latt}=\{ n \in \N \,|\, \exists L\subset \R^n \text{ lattice with } \Delta_n^{\max}=\Delta_{\cP_L} \}\, . 
\end{equation}
\item The set ${\rm OPT}_{\rm Per}$ is the set of dimensions $n\in \N$ such that the maximal
sphere packing density in dimension $n$ is realized by a periodic set $\Sigma$,
\begin{equation}\label{OPTPer}
{\rm OPT}_{\rm Per}=\{ n \in \N \,|\, \exists \Sigma \subset \R^n \text{ periodic set with } \Delta_n^{\max}=\Delta_{\cP_\Sigma} \}\, . 
\end{equation}
\item The set ${\rm OPT}_{\rm Per,N}$ is the set of dimensions $n\in \N$ such that the maximal
sphere packing density in dimension $n$ is realized by a periodic set $\Sigma$ of bounded size $\sigma(\Sigma)\leq N$,
\begin{equation}\label{OPTPerN}
{\rm OPT}_{{\rm Per} \leq N}=\{ n \in \N \,|\, \exists \Sigma \subset \R^n \text{ periodic set with } \sigma(\Sigma)\leq N \text{ and } 
\Delta_n^{\max}=\Delta_{\cP_\Sigma} \}\, .
\end{equation}
\item For a given total recursive function $F:\N\to \N$, the set ${\rm OPT}_{\rm Per,F}$ is the set of dimensions $n\in \N$ such that the maximal
sphere packing density in dimension $n$ is realized by a periodic set $\Sigma$ of bounded size $\sigma(\Sigma)\leq F(n)$,
\begin{equation}\label{OPTPerF}
{\rm OPT}_{{\rm Per}_F}=\{ n \in \N \,|\, \exists \Sigma \subset \R^n \text{ periodic set with } \sigma(\Sigma)\leq F(n) \text{ and } 
\Delta_n^{\max}=\Delta_{\cP_\Sigma} \}\, .
\end{equation}
\end{itemize}
\end{defn}

\smallskip

These sets are related by inclusions
$$ {\rm OPT}_{\rm Latt} \subseteq {\rm OPT}_{{\rm Per} \leq N} \stackrel{N\leq N'}{\subseteq} {\rm OPT}_{{\rm Per} \leq N'} \subseteq  {\rm OPT}_{\rm Per} \, , $$
$$  {\rm OPT}_{\rm Latt} \subseteq  {\rm OPT}_{{\rm Per}_F}  \subseteq  {\rm OPT}_{\rm Per} \, . $$
They are non-empty, since $\{ 1,2,3,8,24 \} \subset {\rm OPT}_{\rm Latt}$, and at least some of the inequalities are strict, since the point $n=10$ is in ${\rm OPT}_{{\rm Per}\leq 40}$ but not in ${\rm OPT}_{\rm Latt}$. 

\smallskip

While the current state of knowledge about these sets does not provide much information on
their structure, the following conjectures reflect the expectations formulated in the 
literature (see for instance \cite{Kumar}).

\begin{conj}\label{conjOPT}
The following conjectures describe the expected behavior of the sets ${\rm OPT}_{\rm Latt}, {\rm OPT}_{\rm Per}$:
\begin{enumerate}
\item[{\bf A}:] The set ${\rm OPT}_{\rm Latt}$ is finite
$$ \# {\rm OPT}_{\rm Latt} < \infty \, . $$
\item[{\bf B}:] (Zassenhaus conjecture) The set ${\rm OPT}_{\rm Per}$ is everything
$$ {\rm OPT}_{\rm Per} = \N \, . $$
\end{enumerate}
\end{conj}

\smallskip

If both conjectures A and B fail, then both ${\rm OPT}_{\rm Latt}$ and ${\rm OPT}_{\rm Per}$ are
infinite sets smaller than the full set $\N$ of positive integers. In this case, it becomes interesting
to ask whether they are computable sets. Similarly, one can 
formulate the same computability question for the intermediate sets  ${\rm OPT}_{{\rm Per} \leq N}$. 

\smallskip

\begin{ques}\label{CompQues}
For the subsets ${\rm OPT}_{\rm Latt}, {\rm OPT}_{\rm Per,N}, {\rm OPT}_{{\rm Per}_F}, {\rm OPT}_{\rm Per}$
one can ask the corresponding computability question:
\begin{enumerate}
\item is ${\rm OPT}_{\rm Latt}$ computable?
\item is ${\rm OPT}_{{\rm Per}\leq N}$ computable?
\item is ${\rm OPT}_{{\rm Per}_F}$ computable?
\item is ${\rm OPT}_{\rm Per}$ computable?
\end{enumerate}
\end{ques}

\smallskip

Note that, even if either or both of Conjectures A and B hold, Question~\ref{CompQues}
remains interesting for the intermediate sets ${\rm OPT}_{{\rm Per}\leq N}$ and 
${\rm OPT}_{{\rm Per}_F}$, and
for either ${\rm OPT}_{\rm Latt}$ or ${\rm OPT}_{\rm Per}$ if only one of
the two conjectures holds. 

\smallskip

A subset of $\N$ is recursively enumerable if there is an algorithm that
enumerates the elements of the set, or equivalently, the set is the
range of a recursive function. In an equivalent formulation, a set of
natural numbers is recursive enumerable if there is an algorithm for
which it is the set of inputs on which the computation halts (the set
is the domain of a recursive function). 
Computability is then equivalent to the condition that both the set
and its complement are recursively enumerable.

\smallskip

In \S \ref{CompOPTsec} we show that all these sets can only be as
non-computable as Kolmogorov complexity, which can be regarded as a
``mild" form of non-computability, since it has good computable upper bounds. 
In other words, they are computable given
an oracle that computes Kolmogorov complexity. This result is obtained using the relation to 
spherical codes, combined with results of
\cite{ManMar2} on the asymptotic bound for spherical codes, and with an extension to spherical
codes of the results of \cite{ManMar1} for $q$-ary codes. The result we obtain
depends on a hypothesis of ``uniform approximability" of the asymptotic bound. 

\medskip
\section{Codes, spherical codes, and asymptotic bounds}\label{SphCodesSec}

In the setting of $q$-ary codes $C\subset \F_q^n$ that are not necessarily linear (unstructured codes), 
the existence of an asymptotic bound in the space of the rate and relative
minimum distance code parameters $(R,\delta)$ was shown in \cite{Man82}. The code
parameters rate and relative minimum distance measure, respectively, how good the encoding
and decoding properties of a code $C$ are. The existence of the asymptotic bound is proved using
certain spoiling operations on codes that alter, in a controlled way the parameters $R$ and $\delta$,
showing that certain regions (controlling quadrangles) in the plane of code parameters can be
densely filled with code points. The resulting planar region outlined by all the controlling quadrangles 
is the undergraph of a function $R=\alpha_q(\delta)$ which defines the asymptotic bound. It is further
shown in \cite{Man12}, \cite{ManMar3} that the asymptotic bound curve $R=\alpha_q(\delta)$ separates
a region of dense code points realized with infinite multiplicity from a region of isolated code points
realized with finite multiplicity. Maximizing the code parameters $R$ and $\delta$ corresponds, respectively,
to optimizing the encoding and decoding properties. There is a tension between these two optimization
questions, as generally improving one of the parameters tends to spoil the other. Thus, codes whose
parameters lie above the asymptotic bound represent codes that have very good properties both
in terms of encoding and of error-correction.  In  \cite{Man12}, a computability question is formulated for the
function $R=\alpha_q(\delta)$ that describes the asymptotic bound. It is shown in \cite{ManMar1}
that this function becomes computable given an oracle that oracle that can list codes by increasing 
Kolmogorov complexity. Given such an oracle, one can obtain an 
iterative (algorithmic) procedure for constructing the asymptotic bound. 
In other words, the asymptotic bound can only be as non-computable as Kolmogorov complexity.
Note that the non-computability of Kolmogorov complexity is regarded as a ``mild" one, in the
sense that Kolmogorov complexity admits good computable approximations from above (given by
compression algorithms), while it does not admit computable approximations from below, due to
the halting problem being unsolvable. 

\smallskip

In the setting considered in this paper, we revisit the results of \cite{ManMar1} on
the asymptotic bound for codes and Kolmogorov complexity and we formulate a
similar result for the case of spherical codes. The existence of an asymptotic bound
for spherical codes was shown in \cite{ManMar2}, again using an argument based on
spoiling operations.

\subsection{Asymptotic bound for $q$-ary codes}
The results of \cite{ManMar1}, relating the asymptotic bound for $q$-ary code to Kolmogorov complexity,
rely on a general result, which we recall here for convenience that provides an explicit algorithm separating
preimages with finite and infinite multiplicity for a total recursive function, given an oracle that produces a 
Kolmogorov ordering (ordering by increasing Kolmogorov complexity) of the source set.
This result is then applied to the case of $q$-ary codes through the characterization of the asymptotic bound as 
the curve separating code points with finite multiplicity from code points with infinite multiplicity.

\smallskip

To phrase the problem more precisely, given the function $f: {\rm Codes}_q \to [0,1]^2\cap \Q^2$ from
codes to code parameters, $f(C):=(R(C),\delta(C))$, one wants to obtain an algorithmic
procedure that inductively constructs preimage sets with finite and infinite multiplicity. If such
a construction is unconditionally possible, then this would show that the asymptotic bound
is a computable function. One can see where the difficulty arises in attempting to construct
such an algorithmic procedure in the following way. Choose an ordering of the code points: 
at step $m$ list the code points in order, up to some growing size $N_m$. One wants to
recursively construct sets $A_m$ and $B_m$, which will be updated at each step of
the algorithm to approximate, respectively,  the set of code points with infinite multiplicity
and the set of code points with finite multiplicity. Initialize the two sets to $B_1=\emptyset$ and
$A_1$ a set of code points up to $N_1$. One then wants to update 
at each step both $A_m$ and $B_m$, so that the first set only contains code points with multiplicity $m$. To
this purpose, going from step $m$ to step $m+1$, new code points listed between $N_m$ and $N_{m+1}$ are first added to 
the set $A_m$, then points (previously in $A_m$ or just added) that do not have an $(m + 1)$-st preimage are moved to the
set $B_{m+1}$. For $m\to \infty$ the sets $A_m$ converge to set of code points of infinite multiplicity and the sets $B_m$ 
converge to set of code points of finite multiplicity. The problem with this procedure lies in the fact that, in order to verify 
which points in the current set $A_m$ have or do not have an $(m+1)$-st preimage, one needs to perform an
infinite search. Thus, the key point is to show that, given an oracle that orders codes by Kolmorogov complexity,
it is possible to turn this into a finite search. This then shows that the asymptotic bound, while it may be non-computable,
its non-computability is only as bad as that of Kolmorogov complexity.  One then proceeds in the following way, which
works quite generally. 

\smallskip
\subsection{Kolmogorov complexity and finite/infinite multiplicities}

Suppose given a total recursive function $f: X \to Y$, and let $\nu_X, \nu_Y$ be structural orderings,
namely bijections $\nu_X: \Z_{\geq 0}\to X$ and $\nu_Y:  \Z_{\geq 0}\to Y$. Let
\begin{equation}\label{nx}
n(x):=\# \{ x'\in X \,|\, \nu_X^{-1}(x')\leq \nu_X^{-1}(x) \, \text{ and } \, f(x')=f(x) \} 
\end{equation}
\begin{equation}\label{Fxfxnx}
F: X \to Y\times \Z_{\geq 0}\ \ \ \   F(x):=(f(x),n(x)) \, . 
\end{equation}
Let 
\begin{equation}\label{XmYm}
X_m:=\{ x \in X \,|\, n(x)=m \} \ \ \text{ and } \ \  Y_m:=f(X_m)\subset Y \, ,
\end{equation}
\begin{equation}\label{YinftyYfin}
Y_\infty :=\cap_{m\geq 1} Y_m \ \ \text{ and } \ \  Y_{fin}:= f(X)\smallsetminus Y_\infty \, . 
\end{equation}

\smallskip

We then have the following result from \cite{ManMar1}, which we summarize here for
later use.

\smallskip

\begin{prop}\label{KandYfininf}
Consider a total recursive function $f: X \to Y$ and structural orderings $\nu_X, \nu_Y$ as above.
The sets $Y_{fin}$ and $Y_\infty$ become computable, given an oracle that orders the 
points of $X$ by increasing Kolmorogov complexity. 
\end{prop}

\smallskip

\proof
The key step in the argument of \cite{ManMar1} is to show that, given $y\in Y_\infty$ and $m\geq 1$,
there is a unique $x_m\in X$ with $y=f(x_m)$ and $n(x_m)=m$, and a constant $c=c(f,u,\nu_X,\nu_Y)>0$,
such that
$$ K_u(x_m) \leq c \, \nu_Y^{-1}(y) \, m  \, \log(\nu_Y^{-1}(y) \, m) \, , $$
where $K_u(x)$ denotes the Kolmogorov complexity of $x\in X$, defined using a Kolmogorov--Schnorr optimal
partial recursive enumeration $u: \Z_{\geq 0} \to X$ so that
$$ K_u(x):= \min \{ k\in \Z_{\geq 0} \,|\, u(k)=x \} \, . $$
Changing $u$ to another such enumeration $v$ changes $K_u$ up to bounded (multiplicative) constants
$$ c_1 \, K_v(x) \leq K_u(x) \leq c_2 K_v(x)\, . $$
The complexity $K_u(x)$ can be equivalently described as the minimal length of a program generating $x$ 
in a Turing machine. 

\smallskip

Then proceed as described above, starting with $B_1=\emptyset$ and with
$A_1$ given by the points $y\in f(X)$ such that $\nu_Y^{-1}(y) \leq N_1$.
At the step $m+1$, with $N_{m+1}> N_m$, list all the points $y\in f(X)$ with $\nu_Y^{-1}(y)\leq N_{m+1}$.
With the help of the oracle, given a point $y$ in this list, one can then search for points $x\in X$ with 
$y=f(x)$ and $n(x)=m+1$ only among those points $x$ with Kolmogorov complexity
bounded by a function of $\nu_Y^{-1}(y)$ as above. This replaces the infinite search with a 
finite search (conditional to the oracle providing the ordering by  Kolmogorov complexity).
Update then the sets so that $A_{m+1}$ contains all the elements $y$ in this list for which there is an $x\in X$ with
$y=f(x)$ and $n(x)=m+1$, and $B_{m+1}$ contains the remaining elements of the list.
In the limit $m\to \infty$ this algorithm separates the sets $Y_\infty$ and $Y_{fin}$ which 
therefore become computable, given an oracle as above. 
\endproof

\smallskip

For the purposes of the present paper, we will need a small variant of this result, which
is provided by the following statement. 

\smallskip

Consider total recursive functions $f: X \to Y$ and 
$g: X \to Z$, and structural orderings $\nu_X,\nu_Y,\nu_Z$. Let
\begin{equation}\label{setNx}
 N(x):=\{ x'\in X \,|\, \nu_X^{-1}(x')\leq \nu_X^{-1}(x) \, \text{ and } \, f(x')=f(x) \} \, ,
\end{equation} 
as in \eqref{nx}. Let $N_Z(x)\subset N(x)$ be a section of $g$ in $N(x)$, namely a subset
of $N(x)$ that does not contain more than one point in each fiber of $g$,
\begin{equation}\label{setNZx}
 \# (N_Z(x)\cap g^{-1}(z))\leq 1\, , \  \forall z\in Z . 
\end{equation} 
Such a set $N_Z(x)$ can be constructed by ordering the points of $N(x)$
according to the structural ordering $\nu_X^{-1}$ and keeping or removing the
next $x'\in N(x)$ according to whether $g(x')$ is different from all the values of
$g$ on the previous elements of $N(x)$ or not. Then set
$$ n_Z(x):= \# N_Z(x)\, ,  $$
and define
\begin{equation}\label{FZx}
 F_Z: X \to Y\times \Z_{\geq 0} \ \ \  F_Z(x)=(f(x),n_Z(x))\, . 
\end{equation} 
As before, define
$$ X_{m,Z}:=\{ x\in X\,|\, n_Z(x)=m \} \ \ \text{ and } \ \ Y_{m,Z}=f(X_{m,Z})\subset Y $$
$$ Y_{\infty,Z}:= \cap_{m\geq 1} Y_{m,Z} \ \  \text{ and } \ \  Y_{fin,Z}:=f(X)\smallsetminus Y_{\infty,Z}\, . $$
The sets $Y_{fin,Z}$ and $Y_{\infty,Z}$ are, respectively, given by
\begin{equation}\label{YZfininf}
 Y_{fin,Z}=\{ y\in f(X) \,|\, \# g(f^{-1}(y))<\infty \} \ \ \text{ and } \ \  
Y_{\infty,Z}=\{ y\in f(X) \,|\, \# g(f^{-1}(y))= \infty \}\, . 
\end{equation}

\begin{cor}\label{fgKandYfininf}
Consider two total recursive functions $f: X \to Y$ and $g: X \to Z$, and structural orderings 
$\nu_X,\nu_Y$. Then the sets $Y_{fin,Z}$ and $Y_{\infty,Z}$ as above become computable
given an oracle that orders the points of $X$ by increasing Kolmorogov complexity.
\end{cor}

\proof The construction of the sets $Y_{fin,Z}$ and $Y_{\infty,Z}$ relies, as in the
case of Proposition~\ref{KandYfininf} on the search for points $x\in X$ with 
$y=f(x)$ and $n(x)=m+1$, which can be made a finite search given an oracle as
in Proposition~\ref{KandYfininf}. Once this search is performed the further test
that determines which of these points belong to the subset $N_Z(x)\subset N(x)$
only involves a finite search within the already constructed $N(x)\cap f^{-1}(y)$. 
\endproof

\smallskip
\subsection{Spherical codes and code parameters}

We now want to modify the setting recalled above to adapt it to the case of spherical codes. We recall the structure of
the set of code parameters and of the asymptotic bound in the case of spherical codes from \cite{ManMar2}. 

\smallskip

\begin{defn}\label{spherecodesparam}
A spherical code in dimension $n$ is a finite set $X$ of points on unit sphere $S^{n-1}\subset \R^n$,
and $\cS\cC$ denotes the set of all spherical codes.
The two code parameters associated to a spherical code $X$ are
\begin{itemize}
\item the {\rm code rate} $R_X =n^{-1}\, \log_2 \# X$;
\item the {\rm minimal angle} $\phi_X$, with the property that $\forall x\neq y\in X$
$$ \langle x, y \rangle \leq \cos\phi_X \, . $$
\end{itemize}
Code points $(R_X,\phi_X)$ are contained in the space $P\cS\cC:=\R_+ \times [0,\pi]$, with
$\bP:\cS\cC\to P\cS\cC$ given by $\bP(X)=(R_X,\phi_X)$. 
Let $A(n,\phi)$ denote the maximal number of points on $S^{n-1}$ with minimal angle $\phi$.
\end{defn}

\smallskip

The space of code parameters of spherical codes is unbounded, since $R_X\to \infty$ for $\phi_X\to 0$.

\smallskip

The optimization questions for spherical codes are:
\begin{itemize}
\item Given $M\in \N$, find a spherical code $X$ with $M$ points, such that minimum angle $\phi_X$
between points of the code is as large as possible.
\item Given an angle $\phi$, find a spherical code $X$ with largest number $M=\# X$ of points 
with angular distance at least $\phi$.
\end{itemize}
As in the case of $q$-ary codes mentioned above, there is a tension between optimizing the two 
code parameters $\phi$ and $R$, see \cite{ErZi}.
For angle separation $\phi_X=\pi/3$ one can view the points of a spherical code as 
contact points for an arrangement of touching non-overlapping equal spheres,
so that, in this case $A(n,\pi/3)$ is the kissing number in dimension $n$.  

\smallskip

Binary codes can be reinterpreted as a special case of spherical codes, by identifying the
set $\{ \pm 1 \}^n$, after a rescaling,  with the $n$-cube centered at the origin of $\R^n$,
inscribed in the sphere $S^{n-1}$. These points define a spherical code $X_C$.
Then the code parameters for an $[n,k,d]_2$-binary code
$C \subset \{ \pm 1 \}^n$ are related to the spherical code parameter by
$$ R(C)=\frac{1}{n} \log_2 \# X_C = R_{X_C} \ \ \text{ and } \ \ \delta(C)=\frac{1-\cos \phi_{X_C}}{2} \, . $$

\smallskip

We will focus on  the following regions of the space $P\cS\cC=\R_+\times [0,\pi]$
of spherical code parameters, as considered in \cite{ManMar2}: 
\begin{itemize}
\item the set of points surrounded by a $2$-ball densely filled by code points
$$ \cU :=\{ P=(R,\phi)\,|\, \exists \epsilon>0 \,: \ B(P,\epsilon)\subset \cA \} \, ; $$ 
\item the asymptotic bound for spherical codes:
$$ \Gamma :=\{ (R=\alpha(\phi),\phi)\,|\, \alpha(\phi)=\sup \{ R\in \R_+\,:\, (R,\phi)\in \cU \} \, \} $$
with $\alpha(\phi)=0$ if $\{ R\in \R_+\,:\, (R,\phi)\in \cU \}=\emptyset$.
\end{itemize}

\smallskip

\begin{rem}\label{GammaAnphi}{\rm
The asymptotic bound $\Gamma=\{ (R,\phi)\,|\, R=\alpha(\phi) \}$ describes the
asymptotic behavior for large $n$ of the ratio $n^{-1} \log_2 A(n,\phi)$, 
with $A(n,\phi)$ the maximal number of code points for a
given minimal angle $\phi$,
\begin{equation}\label{limAnphi}
 \lim_{n\to \infty} \frac{\log_2 A(n,\phi)}{n} =\alpha(\phi)\, . 
\end{equation} 
}\end{rem}

\smallskip

There are new phenomena that happen in the case of spherical codes, with respect to the
case of the asymptotic bound for $q$-ary (and in particular for binary) codes. The region $\cU$ does 
not coincide with the set of accumulation points of the set $\bP(\cS\cP)\subset P\cS\cP$ of code points, and the asymptotic bound is the boundary
of the region $\cU$ (rather than of the set of accumulation points). This is a consequence of the fact that spherical codes admit continuous
parameters unlike binary (or $q$-ary) codes, see \cite{ManMar2} for a more detailed discussion. Note also
that the asymptotic bound is only non-trivial in the ``small angle region"  $0\leq \phi \leq \pi/2$, while
the ``large angle region" $\pi/2 < \phi \leq \pi$ lies entirely above $\Gamma$. This property follows from the
{\em Rankin bound} for spherical codes: for $\pi/2 < \phi \leq \pi$
$$ A(n,\phi) \leq (\cos\phi -1)/\cos\phi \, . $$
This implies that code points of spherical codes of dimension $n$ lie below the curve
$$ R=\frac{1}{n} \log_2 (\min \{ n+1, \frac{\cos\phi-1}{\cos\phi} \}) $$
and for $n\to \infty$ one has $\frac{\log_2 A(n,\phi)}{n} \to 0$ for $\pi/2\leq \phi \leq \pi$. In the
small angle region, instead of the Rankin bound one can use the {\em Kabatiansky--Levenshtein bound}, which for
large $n\to \infty$ gives
$$ R\leq \frac{\log_2 A(n,\phi)}{n} \leq  \frac{1+\sin\phi}{2\sin\phi} \log_2 ( \frac{1+\sin\phi}{2\sin\phi} )
- \frac{1-\sin\phi}{2\sin\phi} \log_2 ( \frac{1-\sin\phi}{2\sin\phi} ) $$
for minimum angle $0\leq \phi \leq \pi/2$. This implies that, for $n\to \infty$, the code points lie in the 
undergraph region
\begin{equation}\label{SunderH}
  \cS :=\{ (R,\phi)\in \R_+\times [0,\pi]\,:\, R\leq H(\phi) \} 
\end{equation}  
where the function $H(\phi)$ is given by
$$ H(\phi)=\frac{1+\sin\phi}{2\sin\phi} \log_2 ( \frac{1+\sin\phi}{2\sin\phi} )
- \frac{1-\sin\phi}{2\sin\phi} \log_2 ( \frac{1-\sin\phi}{2\sin\phi} )\, . $$
This function is unbounded for $\phi\to 0$.

\smallskip

The region $\cU$ in the space of code parameters of spherical codes, and the
asymptotic bound $\Gamma$ are contained in the undergraph $\cS$ of \eqref{SunderH}
(see Lemma~2.9.1 of \cite{ManMar2}).

\medskip
\subsection{Jammed codes}

A deformation of a spherical code is a continuous path in the configuration space
of points on the sphere, with minimum distance bounded below by the value it takes
on the initial configuration.
A code is jammed if the only such deformations are global $SO(n)$-rotations.
In other words, a code is jammed if 
one cannot improve the minimum distance through a continuous motion of
the code points.
A spherical code is infinitesimally jammed if every infinitesimal deformation is an
infinitesimal rotation. An infinitesimally jammed code is jammed (\cite{Conn}, \cite{RoWhi}), 
but (unlike jamming in Euclidean spaces) the converse may not necessarily hold, 
\cite{CJKT}. 

\smallskip

In the space $\cS\cC_n$ of spherical codes $X\subset S^{n-1}$
for a fixed dimension $n$, it is useful to distinguish between the set of rigid (or jammed)
spherical codes $\cS\cC_{n,J}$ and the complementary space of non-jammed codes
$\cS\cC_{n,J}^c$.

\smallskip

\begin{lem}\label{jamcompute}
The set $\cS\cC_{n,J,\bar\Q}$ of spherical codes $X\subset S^{n-1}$ where
code points have algebraic coordinates is decidable.
\end{lem}

\proof
There are efficient linear programming algorithms to test whether a spherical
code is infinitesimally jammed (see Section~2 of \cite{CJKT}). As observed
in \cite{CJKT}, it is possible to determine algorithmically whether a spherical 
code is jammed (even if not infinitesimally jammed), in the case of codes
whose points have algebraic coordinates, using the characterization of
jamming in Proposition~3.2 of \cite{RoWhi} and quantifier elimination for 
the first-order theory of the real numbers of \cite{Tarski}.
Although that is not an efficient algorithm, it shows that the set $ \cS\cC_{n,J,\bar\Q}$
is decidable.
\endproof

\medskip

\section{Computability on metric spaces}

In the case of binary and $q$-ary codes considered in \cite{Man82}, \cite{Man12}, 
\cite{ManMar3}, \cite{ManMar1}, the set of codes in any given dimension $n$ is
a discrete set, given by the subsets of the $q$-ary cube $\F_q^n$. Thus, the
set of codes is countable, and as discussed in \cite{ManMar1} it is in fact a
computable set. Similarly, the set of code parameters $\Q\cap [0,1]^2$ is
also an enumerable set. Therefore, one can consider, as in \cite{ManMar1},
the function $P: \cC_q \to \Q\cap [0,1]^2$ that assigns to a $q$-ary code $C$
its code parameters $(R_C,\delta_C)$ and apply Proposition~\ref{KandYfininf}.
This cannot be directly extended to the case of spherical codes, because 
spherical codes $X\subset S^{n-1}$ have continuous parameters, hence
the ordinary notion of computability cannot be directly applied. However, an 
appropriate notion that allows for continuous parameters was developed for
closed subsets of Euclidean spaces and more generally of metric spaces,
\cite{BraWei}, \cite{BraPre}, \cite{IljKir} and can be used in the setting of 
spherical codes. We recall here the main notions of relevance to our purposes. 

\smallskip

In the classical setting of the theory of computation, a subset $A\subset \N$ (or of
another countable set) is recursive (computable) if there is an algorithm that determines whether 
an element $n\in \N$ belongs to $A$, and recursively enumerable if there is an
algorithm that lists the elements of $A$ (an algorithm that has $A$ as the set of inputs 
on which the program stops). Similarly, $A$ is co-recursively enumerable if the
complement $A^c$ is recursively enumerable. 

\smallskip

For our purposes we define computability for metric space and
continuous functions between metric spaces in the following way,
following \cite{IljKir}.

\begin{defn}\label{computMd}
Let $(M,d)$ be a metric space. A non-empty open set $U\subseteq M$  
is recursively enumerable if there is a computable sequence $\{ x _k \}_{k\in \N} \subset M$
and a computable sequence $\{ r_k \}_{k\in \N}$ in $\R^*_+$ such that
$$ U =\bigcup_{k\in \N} B_d(x_k, r_k)\, , $$
where $B_d(x_k, r_k)$ are the open metric balls in $(M,d)$ centered at $x_k$ of radius $r_k$.
A non-empty closed subset $S\subseteq M$ is recursively enumerable if there is a computable
sequence $\{ x _k \}_{k\in \N} \subset M$ that is dense in $S$. A closed or open
subset $A\subseteq M$ is computable if both $A$ and $M\smallsetminus A$ are
recursively enumerable (with the empty set assumed to be recursively enumerable).
We say that a closed or open subset $A\subseteq M$ is oracle-computable if the
above holds with the sequence $\{ x _k \}_{k\in \N} \subset M$ being computable
given an oracle that can solve a specified (undecidable) problem. 
\end{defn}

\smallskip

\begin{defn}\label{computMd2}
A computable metric space is a triple $(M,d,\mu)$ of a metric space $(M,d)$
and a sequence $\mu: \N \to M$ that is dense in $M$ and such that the
function $d_\mu: \N^2 \to \R$ with $d_\mu(i,j)=d(\mu_i, \mu_j)$
is a computable functions. 
\end{defn} 

\smallskip

In order to have a good notion of computable functions in this setting of
computable metric spaces, one proceeds as follows, \cite{IljKir}.

\smallskip

\begin{defn}\label{FcomputMd}
In a computable metric space $(M,d,\mu)$ a Cauchy name for a point $x\in M$
is a function $p:\N \to \N$ such that $d(x,\mu_{p(k)})< 2^{-k}$, for all $k\in \N$.
Cauchy names determine a partially defined function $\delta_X: {\rm Dom}(\delta_X)\subset \N^\N \to X$
with $\delta_X(p)=x$ iff $p$ is a Cauchy name for $x$. A function $f:(M,d,\mu)\to (M',d',\mu')$
is computable if there is a computable function $\Phi_f:  {\rm Dom}(\Phi_f)\subset \N^\N \to \N^\N$
that maps a Cauchy name for $x$ to a Cauchy name for $f(x)$.
\end{defn}

\smallskip

\begin{rem}\label{mixcompute}{\rm
If $S$ is a subset of a computable metric space $(M,d,\mu_M)$ that is a finite disjoint
union $S=\sqcup_{i=1}^N A_i \sqcup \sqcup_{j=1}^M U_j$ of closed sets $A_i$ and open
sets $U_j$ we say that $S$ is computable if all the $A_i$ and the $U_j$ are computable
in the sense of Definition~\ref{computMd}, with the $x_k\in \mu_M$.
}\end{rem}

\smallskip

We then obtain the following version of Proposition~\ref{KandYfininf} and Corollary~\ref{fgKandYfininf}
in the setting of computable metric spaces.

\begin{thm}\label{YinffinMd}
Let $(X, d_X, \mu_X)$, $(Y,d_Y, \mu_Y)$ and $(Z, d_Z,\mu_Z)$ be computable metric spaces,
with computable functions $f: X\to Y$ and $g: X \to Z$, in the sense of Definition~\ref{FcomputMd},
with the property that $f(\mu_X)\subset\mu_Y$ and $g(\mu_X)\subset \mu_Z$. 
Let $\nu_X$, $\nu_Y$ be structural orderings on the sets $\mu_X$ and $\mu_Y$ respectively. 
Let $\mu_{Y,fin,Z}:=\mu_Y \cap Y_{fin,Z}$ and $\mu_{Y,\infty,Z}:=\mu_Y \cap Y_{\infty,Z}$, with 
$Y_{fin,Z}$ and $Y_{\infty,Z}$ defined as in \eqref{YZfininf}, and let $\bar \mu_{Y,fin,Z}$ and
$\bar \mu_{Y,\infty,Z}$ be the closures in $(Y,d_Y, \mu_Y)$ of these two sets. 
The closed subsets $\bar \mu_{Y,fin,Z}$ and
$\bar \mu_{Y,\infty,Z}$ of $(Y,d_Y, \mu_Y)$ are computable, given the existence of
an oracle that orders the points of $\mu_X$ by increasing Kolmogorov complexity.
\end{thm}

\proof If the computable function $f: X\to Y$
satisfies $f: \mu_X \to \mu_Y$, in particular it gives a computable function in the usual sense
between these countable computable sets, which agrees with the function $D_f$ (with the points
of $\mu_X$ and $\mu_Y$ as tautological Cauchy names of themselves). The same applies to
the computable function $g: X \to Z$ under the same hypotheses. We can then apply 
Corollary~\ref{fgKandYfininf} and obtain that the sets $\mu_{Y,fin,Z}$ and $\mu_{Y,\infty,Z}$ are
computable given an oracle as stated. These are then computable dense subsets of 
$\bar \mu_{Y,fin,Z}$ and $\bar \mu_{Y,\infty,Z}$, respectively, hence the latter are computable
closed subsets of $(Y,d_Y, \mu_Y)$, in the sense of Definition~\ref{computMd}.
\endproof

\smallskip

The following results will also be needed in Section~\ref{CompOPTsec} below.

\smallskip

\begin{lem}\label{fNMd}
Suppose given a computable function $f: \N \to M$, where $(M,d,\mu)$ is 
a computable metric space, and an open or closed computable subset
$A\subset M$. The preimage $f^{-1}(A)\subset \N$
is a computable set in the usual sense. If $A$ is oracle-computable then
so is $f^{-1}(A)$.
\end{lem}

\proof Given computable metric spaces $(M,d,\mu)$ and $(M',d',\mu')$ and
a computable function $f: M\to M'$, 
the preimage operation $A\mapsto f^{-1}(A)$ for $A\subset M'$ open or closed 
is computable (see Theorem~3.30 of \cite{Collins}), and the preimage $f^{-1}(A)$ 
with the metric induced from $d$ is a computable metric space. 
We consider on $\N$ the metric $d(n,m)=1$ for $n\neq m$ and zero otherwise,
which generates the discrete topology.
The computability of $f^{-1}(A)$ in the sense of Definition~\ref {computMd}
in this case means that there is a computable enumeration of the set
$f^{-1}(A)$ hence it agrees with computability in the usual sense. 
\endproof

\smallskip

Moreover, we have the following result from \cite{Brattka}, which will be
useful in Corollary~\ref{KGammaeps} below.
Recall from \cite{Brattka} that a computable metric space $(M,d,\mu)$
is effectively locally connected, if given any $x\in M$ and any open ball
$B_d(x,r)$, it is possible to effectively (through a computable function)
find a connected open set $U$ with $x\in U$ and $U\subseteq B(x,r)$.

\begin{lem}\label{graphComp}
Let $f:M \to M'$ be a function between computable metric spaces
$(M,d,\mu)$ and $(M',d',\mu')$, where $M$ is effectively locally connected. 
Then $f$ is computable, in the
sense of Definition~\ref{FcomputMd} iff the graph $\Gamma(f)\subset M\times M'$
is a computable subset.
\end{lem}

\medskip
\subsection{Metric space of spherical codes}\label{SecMetSC}

For a fixed $n\in \N$, let $\cS\cC_n$ be the set of all spherical codes $X\subset S^{n-1}$.
Each such $X$ is a finite unordered collection of points on the sphere $S^{n-1}$, hence the set
$\cS\cC_n$ can be identified with the configuration space of points
$$ {\rm Config}(S^{n-1})=\bigsqcup_{N\geq 1} {\rm Config}_N(S^{n-1})\, , $$
$$ {\rm Config}_N(S^{n-1})=( (S^{n-1})^N \smallsetminus \Delta_N )/S_n \, , $$
where $\Delta_N$ is the locus of the diagonals (the $N$-tuples where some
of the points coincide) and the symmetric group $S_n$ acts by permutations. 
The ${\rm Config}_N(S^{n-1})$ are smooth non-compact manifolds that are
embedded as the dense open set of non-singular points in the orbifolds
$(S^{n-1})^N/S_n$. The metric 
$$ d_{n,N}(X,Y)=\frac{1}{N}\sum_{i=1}^N d_{S^{n-1}}(x_i,y_i)\, , $$
with $d_{S^{n-1}}(x_i,y_i)$ the round metric on $S^{n-1}$ normalized to
have diameter $1$, gives a metric on $(S^{n-1})^N$ invariant under
permutations, which induces a metric on the orbifold $(S^{n-1})^N/S_n$ and
by restriction on the configuration space ${\rm Config}_N(S^{n-1})$.
This in turn determines a metric on ${\rm Config}(S^{n-1})$ by
$$ d_n(X,Y)=1 \ \ \text{ if } X \in {\rm Config}_N(S^{n-1})\, , Y\in {\rm Config}_M(S^{n-1})\ \ 
\text{ with } N\neq M $$
$$ d_n(X,Y)=d_{n,N}(X,Y) \ \ \text{ if } X,Y \in {\rm Config}_N(S^{n-1}) \, . $$
One can then consider the space of all spherical codes,
$$ \cS\cC = \bigsqcup_n \cS\cC_n = \bigsqcup_n {\rm Config}(S^{n-1})=\bigsqcup_{n,N} 
{\rm Config}_N(S^{n-1})\, , $$
with the induced metric $d(X,Y)=d_n(X,Y)$ for $X,Y\subset S^{n-1}$ and
$d(X,Y)=1$ otherwise. This metric extends in the same form to the disjoint
union of the orbifolds $\cO:=\sqcup_{n,N} (S^{n-1})^N/S_N$, with $\cS\cC\subset \cO$
a dense open set. 

\smallskip

We can further identify spherical codes in $\cS\cC_n$ that are related by a global
$SO(n)$-isometry of $S^{n-1}$, as they only differ by a rigid motion of the
ambient sphere. The metric described above is invariant hence it descends to
the quotient. We still denote the resulting spaces by $\cS\cC$ and $\cO$ 
for simplicity of notation.

\smallskip

\begin{prop}\label{computSC}
The metric space $(\cO,d)$ introduced above is
a computable metric space in the sense of Definitions~\ref{computMd}
and \ref{computMd2}, 
and the metric space $(\cS\cC,d)$ of spherical codes is a computable
open subset in the sense of Definition~\ref{computMd}.
\end{prop}

\proof Consider the $S_n$-invariant subset of $(S^{n-1})^N$ given by
the $N$-tuples of points $(x_1,\ldots, x_N)$ such that the angular
coordinates $(\varphi_1, \ldots, \varphi_{n-1})$ of the points
$x_i$, in are all rational, 
with the first angle
measuring the minimal distance. 
This is a computable dense subset of the classes of $(S^{n-1})^N/S_n$
up to global $SO(n)$-isometries of $S^{n-1}$.
Let $\mu_n$ be a computable enumeration of this subset and let $\mu_\cO$
be the resulting diagonal enumeration of a computable dense subset of $\cO$.
The restriction to $\cS\cC$ is a dense computable set $\mu_{\cS\cC}=\mu_{\cO}\cap \cS\cC$,
and so is $\mu_{\cS\cC^c}=\mu_{\cO}\cap \cS\cC^c$ in the closed complement
$\cS\cC^c=\cO\smallsetminus \cS\cC$. The
metric balls $B_d(x,r)$ with $x\in \mu_{\cS\cC}$ and $r\in \Q^*_+$ exhaust 
$\cS\cC$, which is therefore a computable open subset. 
\endproof

\smallskip

\begin{prop}\label{PcomputSC}
The space $P\cS\cC =\R_+ \times [0,\pi]$ of code parameters of spherical codes is
a computable metric space and the function $\bP: \cS\cC\to P\cS\cC$ mapping
a code $X$ to its code parameters $\bP(X)=(R_X,\phi_X)$ is a computable
function between computable metric spaces. The dimension function
$D: \cS\cC \to \N$ that assigns to a code its dimension
$D(X\subset S^n)= n$ is also a computable function. 
\end{prop}

\proof
Consider in $P\cS\cC =\R_+ \times [0,\pi]$ the computable subset of points
$P=(R,\phi)$ with $R\in \Q \log_2\N$ and $\phi\in \Q$, 
with an enumeration $\mu_{P\cS\cC}$. Given a
point $X\in \cS\cC$ and a Cauchy name $p_X$ with $\mu_{\cS\cC}(p_X(k))=
X_k$ a code with points with rational angular coordinates, with distance at
most $2^{-k}$ from $X$. The code points
$\bP(X_k)$ give an approximating sequence in $P\cS\cC$ to
$\bP(X)$, with $\bP(X_k)\in \mu_{P\cS\cC}$.  
Since the distance between $X_k$ and $X$ in $\cS\cC$ is
bounded by $2^{-k}$, we have $X_k,X\in {\rm Config}_N(S^{n-1})$,
with $N=\# X_k=\# X$, so that 
$d_{P\cS\cC}(\bP(X_k),\bP(X))=| \phi_{X_k}-\phi_X |$. Thus,
if $d_{\cS\cC}(X_k,X)< 2^{-k}$ we also have that
$\bP(X_k)$ is a Cauchy name for $\bP(X)$. This determines the
computable function $\Phi_P$ so $P$ is computable in the sense of
Definition~\ref{FcomputMd}. The dimension function $D$ is also
computable since it is constant on all the components 
${\rm Config}(S^{n-1})$.
\endproof

\smallskip

\begin{rem}\label{jamrem}{\rm
For fixed $n$ and $N$, we can consider the minimum angle
function $\phi: X\mapsto \phi_X$ for codes in ${\rm Config}_N(S^{n-1})$,
considered modulo global $SO(n)$-rotations. The isolated 
local maxima of $\phi$ are the jammed codes $X\in \cS\cC_{n,J}$ 
with $\# X=N$. The code points 
$(R=\frac{\log_2 N}{n}, \phi)$ that are local maxima of $\phi$, 
for which all codes $X\subset S^{n-1}$ with $\bP(X)=(R,\phi)$
are jammed codes, are the code points 
with finite multiplicity $\# \bP^{-1}(R,\phi)<\infty$ in ${\rm Config}_N(S^{n-1})$.}
\end{rem}

\medskip
\subsection{Kolmogorov complexity and the asymptotic bound for spherical codes}\label{KolmogSec}

We recall the following result from  \cite{ManMar2}, characterizing the region
below the asymptotic bound.

\smallskip

\begin{prop}\label{infmultipl} {\rm (Theorem~2.10.1 of \cite{ManMar2})}
A code point $P=(R,\phi)\notin \Gamma$ is contained in the region $\cU$ if and
only if it there exists a sequence $X_k$ of spherical codes $X_k\subset S^{n_k-1}$
with $n_k\to \infty$ and $(R_{X_k},\phi_{X_k})=(R,\phi)$.
\end{prop}

\smallskip

In view of Corollary~\ref{fgKandYfininf} above, and Theorem~\ref{KcomputeGamma} below, 
it is useful to rephrase the above statement in the following way.

\begin{cor}\label{infmultDim}
Let $\cS\cC$ denote the set of all spherical codes and let $D: \cS\cC \to \N$ be
the function that assigns to a code its dimension
$$ D: X\subset S^n \mapsto n \, . $$
Let $P\cS\cC:=\R_+ \times [0,\pi]$ be the set of code parameters of spherical codes
and let $\bP: \cS\cC \to P\cS\cC$ be the function that maps a spherical code to
its code parameters $\bP(X)=(R_X,\phi_X)$
Then the region $\cU$ in the space of code parameters of spherical codes
is given by
$$ \cU = P\cS\cC_{\infty,\N} =\{ (R,\phi)\in P\cS\cC \, |\, \# D(\bP^{-1}(R,\phi)) = \infty \} \, . $$
\end{cor}

\smallskip

We then obtain the following result on the computability properties
of the asymptotic bound for spherical codes.

\begin{thm}\label{KcomputeGamma}
The asymptotic bound $\Gamma$ for spherical codes is
oracle-computable given an oracle that orders the points of $\mu_{\cS\cC}$
by Kolmogorov complexity. The sets $\bar \mu_{P\cS\cC,fin/\infty,\N}$
are also oracle-computable with such an oracle.
\end{thm}

\proof The statement follows from Theorem~\ref{YinffinMd}, applied
to the computable metric spaces $(\cS\cC, d_{\cS\cC}, \mu_{\cS\cC})$
and $(P\cS\cC, d_{P\cS\cC}, \mu_{P\cS\cC})$ with the computable functions
$\bP: \cS\cC\to P\cS\cC$ and $D: \cS\cC\to \N$, as in Proposition~\ref{PcomputSC}. 
Note that these computable functions satisfy the hypotheses of Theorem~\ref{YinffinMd},
since $\mu_{\cS\cC}$ consists of those configurations of points on spheres 
with rational angles and minimum distance (up to a global rotation), hence
the image of $\mu_{\cS\cC}$ under $\bP: \cS\cC\to P\cS\cC$ is contained in
$\mu_{P\cS\cC}=\Q\log_2\N \times \Q$. Moreover, we have 
$\mu_{P\cS\cC,\infty,\N}=\mu_{P\cS\cC}\cap \cU$ by Corollary~\ref{infmultDim},
hence $\bar\mu_{P\cS\cC,\infty,\N}=\bar\cU=\cU\cup \Gamma$. 
Theorem~\ref{YinffinMd} gives that the sets  $\bar\mu_{P\cS\cC,\infty,\N}$ and
$\bar\mu_{P\cS\cC,{\rm fin},\N}$ are computable given the oracle, and 
the asymptotic bound $\Gamma$ is then also an oracle-computable set
as $\bar\mu_{P\cS\cC,\infty,\N}\cap \bar\mu_{P\cS\cC,{\rm fin},\N}$ (see
Theorem~2.27 of \cite{Collins}).
\endproof

\smallskip

For a fixed dimension $n$, the code points with finite multiplicity are coming
from the presence of rigid jammed codes in $S^{n-1}$. Jammed codes can
become unjammed in higher dimension, when the given sphere is embedded
in a higher dimensional one. These unjammed codes correspond to code
points with a smaller value of $R$ (see the discussion of spoiling operations 
on spherical codes in \cite{ManMar2}). These point may lie in $\cU$ or
in the region of accumulation points of $\bP(\cS\cC)$. The use of the
function $D: \cS\cC\to \N$ and the sets $\mu_{P\cS\cC,\infty,\N}$ 
instead of just the sets $\mu_{P\cS\cC,\infty}$ separates these
two possibilities and focuses on the case of points in $\cU$.

\medskip
\subsection{Oracle computability and sublevel sets} 

The following variant of the above statement will be needed in Section~\ref{CompOPTsec}.

\begin{cor}\label{KGammaeps}
The sets $\Gamma_\epsilon=\{ (R,\phi)\,|\, R\geq \alpha(\phi)-\epsilon \}$ in
the computable metric space $P\cS\cC$ are oracle-computable closed subsets,
given an oracle that orders the points of $\mu_{\cS\cC}$ by Kolmogorov
complexity.
\end{cor}

\proof Assuming an oracle as above, the asymptotic bound $\Gamma$ 
is computable (Theorem~\ref{KcomputeGamma}). By Lemma~\ref{graphComp}
we know that $\Gamma$ computable is equivalent to the function $\alpha: [0,\pi]\to \R_+$,
$R=\alpha(\phi)$ with $\Gamma=\Gamma(\alpha)$ being computable, since
the source space, with its structure of computable metric space induced
from $P\cS\cC$ is effectively locally connected. For a fixed $\epsilon>0$ the
function $\alpha_\epsilon(\phi)=\alpha(\phi)-\epsilon$ is also a computable
function, hence its graph $\Gamma(\alpha_\epsilon)$ is a computable closed
subset of $P\cS\cC$. The region $\Gamma_\epsilon$ is given by
$$ \Gamma_\epsilon = P\cS\cC_{fin}
\cup \Gamma(\alpha) \cup (\Gamma(\alpha)^c \cap \Gamma(\alpha_\epsilon)^c \cap P\cS\cC_{\infty})
\cup \Gamma(\alpha_\epsilon)\, , $$
where $\Gamma(\alpha)^c$ and $\Gamma(\alpha_\epsilon)^c$ are the complements in $P\cS\cC$.
In general an intersection of computable sets in computable metric spaces need not necessarily be
computable (an example with computable compact sets is discussed in  \cite{CebKre}). However,
in this case if $\Gamma(\alpha)^c$ and $\Gamma(\alpha_\epsilon)^c$ 
are computable open sets in the sense of Definition~\ref{computMd}, then the open set
$\Gamma(\alpha)^c \cap \Gamma(\alpha_\epsilon)^c \cap P\cS\cC_{\infty}$, which
consists of the points of $P\cS\cC$ in between the curves $\Gamma(\alpha)$ and  
$\Gamma(\alpha_\epsilon)$ is also exhausted by the balls $B_{d_{P\cS\cC}}(x_k,r_k)$
with centers in $\mu_{P\cS\cC}$ contained in this region hence it is a computable set,
since the intersections of $\mu_{P\cS\cC}$ with each side of $\Gamma(\alpha)$ and  
of $\Gamma(\alpha_\epsilon)$ are computable by the computability of these two sets,
hence $\Gamma_\epsilon$ is also computable, in the sense of Remark~\ref{mixcompute}.
\endproof

\smallskip

Note that also in general the level sets and sublevel sets or suplevel sets of
computable functions in computable metric spaces are not always necessarily
computable, see  \cite{CebKre} for a more detailed discussion.
For the computability of finite intersections of computable open sets, see also Theorem~3.27
of \cite{Collins}.

\medskip
\section{Spherical codes and sphere packings}\label{SphCodesSec}

\subsection{Code and packing densities}
Given a spherical code $X\subset S^{n-1}$, the code density $\Delta_X$ is defined
as the fraction of the $(n-1)$-dimensional area of the sphere $S^{n-1}$ that is
covered by $\# X$ spherical caps of angular radius $\phi_X/2$, that is,
\begin{equation}\label{codeDelta}
\Delta_X = \frac{\# X \cdot S(n,\phi_X)}{S_n} \, ,
\end{equation}
where $S_n=n \pi^{n/2}/\Gamma(1+ n/2)$ is the $(n-1)$-dimensional area of the sphere $S^{n-1}$
and
$$ S(n,\phi)=S_{n-1} \int_0^{\phi/2} \sin^{n-2}(x) dx $$
is the $(n-1)$-dimensional area of a spherical cap of angular radius $\phi/2$. 

\smallskip

\begin{rem}\label{DeltaPreim}{\rm Note that the spherical code density $\Delta_X$
of \eqref{codeDelta} only depends on $X$ through its code parameters
$\bP(X)=(R_X,\phi_X)$, with $\# X =2^{n R_X}$, hence it is constant on the set of
spherical codes that are preimages of the same code point.}
\end{rem}

\smallskip

The maximal possible density for spherical codes in $S^{n-1}$ with
minimal angle $\phi$ is
\begin{equation}\label{maxdensephi}
 \Delta(n,\phi)=  A(n,\phi) \frac{S(n,\phi)}{S_n} \, . 
\end{equation}
One defines
\begin{equation}\label{Deltacoden}
\Delta^{codes}_n:= \lim_{\phi\to 0} \Delta(n,\phi)\, .
\end{equation}
This limit density is related to the maximal sphere packing density by 
\begin{equation}\label{spherecodeDelta}
\Delta^{codes}_n = \Delta^{\rm max}_{n-1} \, .
\end{equation}

\smallskip

A family $X_k$ of spherical codes $X_k\subset S^{n-1}$ with
$\phi_{X_k}\to 0$ as $k\to \infty$ is an asymptotically optimal family
if 
$$ \lim_{k\to \infty} \frac{\# X_k}{A(n,\phi_k)} =1\ \ \ \text{ or equivalently } \ \ 
\lim_{k\to \infty} \frac{\Delta_{X_k}}{\Delta(n,\phi_k)} =1 \, . $$

\smallskip
\subsection{Wrapped spherical codes}\label{WrappedSec}

Given a sphere packing $\cP\subset \R^n$, there are different possible ways
of associating to it a family of spherical codes. Some of these methods
are summarized in \S 3.1 of \cite{ManMar2}. Here we consider only the
construction based on wrapped spherical codes, \cite{HaZe}.

\smallskip

Suppose given an arbitrary sphere packing $\cP$ in $\R^{n-1}$ with packing radius $d=\ell/2$ with
$\ell$ the minimum distance between sphere centers, and with density $\Delta_\cP$. Let $S_\cP$
denote the corresponding set of sphere centers in $\R^{n-1}$. We consider modifications of 
the packing $\cP$ that scale the packing radius $d$ while maintaining the same density
$\Delta_\cP$. In the case of lattice and periodic packings, this is achieved by scaling in the
same way the minimum length and the covolume in \eqref{centerdense} and \eqref{centerdenseN}. 
In this way we can take a limit for $d\to 0$, while maintaining $\Delta_\cP$ fixed.

\smallskip

The wrapped spherical  codes construction of \cite{HaZe} is obtained as follows. 
The latitude of a point $x \in S^{n-1}$ is the angle $\vartheta(x)\in [-\pi/2,\pi/2]$ from a fixed equator of $S^{n-1}$.
A choice of a set of latitude angles $\vartheta=(\vartheta_0,\ldots,\vartheta_N)$ with 
$-\pi/2=\vartheta_0<  \cdots < \vartheta_N =\pi/2$, separates the sphere $S^{n-1}$
into a collection of annuli $A_i$ corresponding to the points with $\vartheta_i\leq \vartheta(x) < \vartheta_{i+1}$.
Let $\phi_i=\vartheta_{i+1}-\vartheta_i$ be the corresponding angular separation of each annulus.
Continuous low distortion maps $f_i: A_i \to U_i\subset \R^{n-1}$ to corresponding annular regions in the
plane are constructed. A spherical code $X_{\cP,\vartheta}$ is then obtained by taking
\begin{equation}\label{XPtheta}
 X_{\cP,\vartheta}=\bigcup_i f_i^{-1}(S_\cP \cap U_i)\smallsetminus B_i \, , 
\end{equation} 
 where $B_i$ is a ``buffer region" of points at distance less than $d$
from the border between the annuli $A_i$ and $A_{i-1}$. The minimal distance 
of points in $X_{\cP,\vartheta}$ is at least $d$. (The buffer region ensures this minimum distance
requirement is met across different annuli.) 

\smallskip

It is shown in \cite{HaZe} that, if the construction above is applied to a family of rescaled
packings $\cP_d$ with $d\to 0$ and with constant density $\Delta_\cP$, where the annuli $A_{i,d}$
with angles $\vartheta_d=(\vartheta_{i,d})_i$ are chosen so that 
$\lim_{d\to 0} \max_i \phi_{i,d} + \frac{d}{\min_i \phi_{i,d}} =0$, 
then the spherical code densities $\Delta_{X_{\cP_d,\vartheta_d}}$ approximate the sphere packing density,
\begin{equation}\label{approxDelta}
\lim_{d\to 0} \Delta_{X_{\cP_d,\vartheta_d}} = \Delta_\cP \, . 
\end{equation}
In particular, if $\cP$ is a sphere packing that realizes the maximal density,
$\Delta_\cP=\Delta^{\rm max}_{n-1}$, the wrapped spherical codes are
an asymptotically optimal family with densities that 
approximate the maximal sphere packing density, 
$$ \lim_{d\to 0} \Delta_{X_{\cP_d,\vartheta_d}} = \Delta^{\rm max}_{n-1}\, . $$

\smallskip

Let $\cS\cC=\cup_n \cS\cC_n$ denote the set of all spherical codes, with
$\cS\cC_n$ the set of spherical codes $X\subset S^{n-1}$. As above, let $\bP: \cS\cC\to \R_+\times [0,\pi]$
be the map that assigns to a spherical code $X$ its code parameters $\bP(X)=(R_X,\phi_X)$.
As above, we let $\cS\cP_{n-1}$ denote the set of all sphere packings $\cP\subset \R^{n-1}$, and
we let $\cS\cP_{n-1}^{\rm max}$ denote the set of sphere packings $\cP\subset \R^{n-1}$ that
realize the maximal density $\Delta_\cP=\Delta^{max}_{n-1}$. 

\smallskip

Consider then a collection of maps $\bP_{n,\vartheta_d}: \cS\cP_{n-1}\to \R_+\times [0,\pi]$ 
that map a sphere packing $\cP\in \cS\cP_{n-1}$ to the code point
$\bP(X_{\cP_d,\vartheta_d})$ of the wrapped spherical code $X_{\cP_d,\vartheta_d}$ associated 
to the packing $\cP_d$ as in \S \ref{WrappedSec} above,
\begin{equation}\label{Pnd}
\bP_{n,\vartheta_d}(\cP)=\bP(X_{\cP_d,\vartheta_d})=(R_{X_{\cP_d,\vartheta_d}},\phi_{X_{\cP_d,\vartheta_d}})\, . 
\end{equation}

\smallskip

\begin{defn}\label{gammadiscr}
Let $\cP\subset \R^{n-1}$ be a sphere packing that is not optimal.
Then the discrepancy of $\cP$ is $1-\gamma$, where
\begin{equation}\label{gamma}
\gamma:= \frac{\Delta_\cP}{\Delta^{\max}_{n-1}} \, . 
\end{equation}
\end{defn}

\smallskip

\begin{prop}\label{XdGamma}
Let $\cP\subset \R^{n-1}$ be a sphere packing of discrepancy $1-\gamma$. 
Then for any $\epsilon>0$ there is a $d_0$ such that, for all $d\leq d_0$
the points $\bP(X_{\cP_d,\vartheta_d})$ are in $\Gamma_{\frac{-\log_2\gamma}{n} +\epsilon}$.
\end{prop}

\proof Since the packing has discrepancy $1-\gamma$, we have for $d\to 0$
$$ R_{X_{\cP_d,\vartheta_d}} \sim \frac{\log_2 A(n,\phi_{X_{\cP_d,\vartheta_d}})}{n} + \frac{\log_2\gamma}{n} \sim \alpha(\phi_{X_{\cP_d,\vartheta_d}}) + \frac{\log_2\gamma}{n} \, . $$
\endproof

\medskip
\section{Oracle computability properties of the OPT sets}\label{CompOPTsec}

First observe that what matters in the computability question for the sets
${\rm OPT}_{\rm Latt}$, ${\rm OPT}_{{\rm Per}\leq N}$, ${\rm OPT}_{\rm Per}$
is the behavior for large $n$, assuming they are infinite sets (see the discussion in \S \ref{OPTsec}). 

\smallskip

\begin{lem}\label{compSPmet}
Let $\cS\cP_{n-1}$ be the set of sphere packings $\cP\subset \R^{n-1}$,
and let $\cS\cP:=\sqcup_n \cS\cP_{n-1}$.
The space $\cS\cP$ of sphere packings is a computable metric space.
\end{lem}

\proof 
As in the construction of wrapped spherical codes of \cite{HaZe}, 
recalled in \S \ref{WrappedSec} above, we partition
the sphere $S^{n-1}$ into a collection of annuli $A(\vartheta)=\{ A_i \}$ with low-distortion maps 
$f_i: A_i \to U_i\subset \R^{n-1}$ to annular regions in $\R^{n-1}$. 
Given the computable metric space structure on $S^{n-1}$ with
$\mu_{S^{n-1}}$ given by an enumeration of the computable subset
of points on the sphere with rational angular coordinates (that is, angles
in $\Q\pi$), we can choose a compatible computable metric space structure
on $\R^{n-1}$ such that annuli $A_i, U_i$ are computable subspaces
and the low-distortion functions $f_i : A_i \to U_i$ are computable
functions in the sense of Definition~\ref{FcomputMd}.
For a sphere packing $\cP\subset \R^{n-1}$ we write 
$$ X_{\cP_i}:=f_i^{-1}(S_\cP\cap U_i)\smallsetminus B_i\, , $$
with $X_{\cP,\vartheta}=\cup_i X_{\cP_i}$ the wrapped code as in \eqref{XPtheta}.
We then define the distance $d(\cP,\cP')$ between two sphere packings
in $\cS\cP$ as 
\begin{equation}\label{dPP}
d_{\cS\cP}(\cP,\cP')=\sup_\vartheta d_{\cS\cC} (X_{\cP,\vartheta} ,X_{\cP',\vartheta}) \, . 
\end{equation}
By the construction of the distance $d_{\cS\cC}$ in \S \ref{SecMetSC},
all the distances $d_{\cS\cC} (X_{\cP,\vartheta} ,X_{\cP',\vartheta})$ are
bounded above by $1$, hence the supremum is bounded. 
We obtain $\mu_{\cS\cP}$ using an enumeration of
a computable dense set of $\vartheta$'s and, for each such $\vartheta$,
the computable dense subsets $f_i(\mu_{\cS\cC}\cap A_i)$ in $U_i$.
\endproof

\smallskip

\begin{lem}\label{tildePcomp}
The functions $\bP_{n,\vartheta_d}: \cS\cP_{n-1}\to P\cS\cC$ of \eqref{Pnd}
are computable functions between computable metric spaces. 
\end{lem}

\proof The choice of metric $d_{\cS\cP}$ on $\cS\cP$ as in \eqref{dPP} and 
the choice of $\mu_{\cS\cP}$ imply that the functions 
$X_{\vartheta_d}: \cP \mapsto X_{\cP,\vartheta_d}$ are computable. 
Combined with Proposition~\ref{PcomputSC} this then ensures that the
composite maps $ \cP \mapsto \bP(X_{\cP,\vartheta_d})$ are also computable
in the sense of Definition~\ref{FcomputMd}. 
\endproof

\smallskip

\begin{prop}\label{compPGammaepsilon}
Consider a computable sequence $\vartheta_{d_k}\to 0$ and the corresponding
computable maps $\bP_{n,\vartheta_{d_k}}: \cS\cP_{n-1}\to P\cS\cC$. For a
fixed oracle-computable closed set $\Gamma_\epsilon=\{ R \geq \alpha(\phi)-\epsilon\}$, the set
\begin{equation}\label{PreimGammaepsilon}
{\mathfrak P}_n:= \bigcap_{k \geq k_0} \bP_{n,\vartheta_{d_k}}^{-1}(\Gamma_\epsilon) \subset \cS\cP_{n-1} 
\end{equation}
is also an oracle-computable set.
\end{prop}

\proof By Lemma~\ref{fNMd} the preimages $\bP_{n,\vartheta_{d_k}}^{-1}(\Gamma_\epsilon)$ are
oracle-computable closed sets, and since countable intersections of closed sets preserves
computability by Theorem~3.27 of \cite{Collins}, so is also the set of \eqref{PreimGammaepsilon}.
\endproof

\smallskip

Let $A\in \{ {\rm Latt}, {\rm Per}\leq N,  {\rm Per}_F, {\rm Per} \}$.
Let $\cS\cP_{n-1}^A$ be the space of sphere packings of
type $A$ and let $\cS\cP_{n-1}^{A,\max}\subset \cS\cP_{n-1}^A$ 
be the subset of those packings that maximize the density among
packings of type $A$. We set
\begin{equation}\label{SPAmax}
\cS\cP^{A,\max}=\bigcup_{n\in \N} \cS\cP_{n-1}^{A,\max}\, .
\end{equation}

\smallskip

The set $\cS\cP^{{\rm Latt},\max}_{n-1}$ can be identified with the
algorithmically computable (Voronoi algorithm) set of vertices 
of the Ryshkov polytope $\cR_n$ \cite{Rysh}.  In a similar way,
for $N$-periodic sets (unions of $N$ translates of a lattice)
one can construct a similar generalized Ryshkov polytope, see
for instance Section~2.3 of \cite{AndKall}. These 
generalized Ryshkov polytopes $\cR_{N,n}$ are less well
understood than the lattice case. The following conjecture
is formulated in Section~2.4 of \cite{AndKall}.

\smallskip

\begin{conj}\label{RNconj}
The generalized Ryshkov polytopes $\cR_{N,n}$ has finitely many
vertices, that can be algorithmically determined. 
\end{conj}

\smallskip

In the case of the generalized Ryshkov polytope $\cR_{2,n}$, an
explicit algorithmic numeration of its vertices is obtained in \cite{AndKall}.
A difficulty that arises in the periodic sets case, which is not present in
the case of lattices, is the occurrence of ``fluid packings"
(see Definition~2 and Theorem~3 of \cite{AndKall}).
In the case $N=2$ the fluid packings are manageable since they
can be described as a union of two lattices that must be
in $\cS\cP^{{\rm Latt},\max}$, but for higher $N$ this becomes more
difficult. 

\smallskip

\begin{prop}\label{CompSPAmax}
Assuming Conjecture~\ref{RNconj} above, 
the sets $\cS\cP^{A,\max}$ are computable for
$A\in \{ {\rm Latt}, {\rm Per}\leq N, {\rm Per}_F, {\rm Per} \}$.
\end{prop}

\proof The existence of a computable enumeration of $\cS\cP^{{\rm Latt},\max}$
follows from the description of the $\cS\cP^{{\rm Latt},\max}_{n-1}$ in terms of
vertices of the Ryshkov polytopes $\cR_n$ and the Voronoi algorithm. Assuming
Conjecture~\ref{RNconj}, the same holds for $\cS\cP^{{\rm Per}\leq N,\max}$
and for $\cS\cP^{{\rm Per}_F,\max}$. The computability of the 
$\cS\cP^{{\rm Per}\leq N,\max}$ in turn can be used to obtain  
a computable enumeration of $\cS\cP^{{\rm Per},\max}$.
\endproof

\smallskip

We then obtain the following result on the computability properties of
the sets ${\rm OPT}_{\rm Latt}$, ${\rm OPT}_{{\rm Per}\leq N}$, 
${\rm OPT}_{\rm Per}$.

\smallskip

\begin{lem}\label{LimAboveGamma}
For $A\in \{ {\rm Latt}, {\rm Per}\leq N, {\rm Per}_F, {\rm Per} \}$ the set
${\rm OPT}_A$ can be described as
$$ {\rm OPT}_A = \{n \in \N \,|\, \cS\cP^{A,\max} \cap {\mathfrak P}_n \neq \emptyset \} \, . $$
\end{lem}

\proof This directly follows from the definitions of ${\mathfrak P}_n$ as in \eqref{PreimGammaepsilon} 
and $\cS\cP^{A,\max} $ as in \eqref{SPAmax}, and from Proposition~\ref{XdGamma}. 
\endproof

\smallskip

\begin{thm}\label{KcompOPT}
The sets ${\rm OPT}_{\rm Latt}$, ${\rm OPT}_{{\rm Per}\leq N}$, 
${\rm OPT}_{\rm Per}$ are oracle-computable given an oracle that
orders the points of $\mu_{\cS\cC}$ by Kolmogorov
complexity.
\end{thm}

\proof Given the computability properties of the sets $\cS\cP^{A,\max}$, 
as in Proposition~\ref{CompSPAmax}, and the oracle computability of
the ${\mathfrak P}_n$ as in Proposition~\ref{compPGammaepsilon}, 
it follows (again Theorem~3.27 of \cite{Collins}) that the sets
$\cS\cP^{A,\max} \cap {\mathfrak P}_n$ are also oracle-computable. 
for $A\in \{ {\rm Latt}, {\rm Per}\leq N, {\rm Per}_F, {\rm Per} \}$, 
by Lemma~\ref{LimAboveGamma}, we can then identify the
${\rm OPT}_A$ as the image under the computable dimension
function $D: \cS\cP \to \N$ of these oracle-computable sets.
It follows
that the sets ${\rm OPT}_A$ are also oracle-computable.
\endproof

\medskip
\subsection{Computability or non-computability?}

Our proof of oracle-computability of the ${\rm OPT}_A$ sets in Theorem~\ref{KcompOPT} leaves
open the question of whether these sets are genuinely computable (even without oracle) or are actually
non-computable. We do not have an answer to this question at this point, but we outline here some heuristic 
reasons for expecting one or the other possibility, and we explain the main difficulties in implementing an
actual proof of computability or of non-computability. 

\smallskip

As already observed by Hales in \cite{Hales} the
decidability of the optimal sphere packing problem in arbitrary dimension is a question about which nothing
definite can be said: the very few existing solutions rely on conditions that are sufficient but very far from
being necessary, so that, in terms of our ${\rm OPT}_A$ sets above, their failure in a given dimension
$n$ would not be able to rule out the possibility that $n\in {\rm OPT}_A$. This presents a main
obstacle towards the construction of an algorithmic procedure that decides whether $n\in {\rm OPT}_A$
is satisfied for a given $n$. 

\smallskip

Proposition~\ref{CompSPAmax} (conditional to Conjecture~\ref{RNconj}) would provide a computable
set of candidates for optimization. However, even given, for instance, an explicit Cohn-Elkies type function
providing an upper bound \cite{CoEl} in every dimension (see \cite{ManMar4}), failure of such a bound 
to be saturated at any of the candidates in dimension $n$ would not be enough to decide the non-belonging
of $n$ to the corresponding ${\rm OPT}_A$ set. This failure of a possible strategy to determine computability, 
however, does not constitute proof of non-computability.

\smallskip

Parts (A) and (B) of Conjecture~\ref{conjOPT} would imply computability. The main evidence for
part (A) is the numerical evidence showing that the optimal lattice-packing density (which is
determined by the Hermite constant in dimension $n$) and the largest density known among
non-lattice packings in the same dimension, wherever one can compute the latter, seems to widen 
as $n$ grows. Support for part (B) comes from the fact that periodic packings approximate
arbitrarily well the maximal density. 

\smallskip

Proving non-computability, on the other hand, amounts to showing that a procedure that
decides whether $n\in {\rm OPT}_A$ for each $n\in \N$ would require the answer to a known
undecidable problem. For example, if {\em any} computing strategy for 
testing $n\in {\rm OPT}_A$ requires determining 
whether a given program halts on the input $n$, the undecidability of the Halting problem
would imply non-computability of ${\rm OPT}_A$. 

\smallskip

Note that the non-computability of Kolmogorov complexity indeed depends on the
undecidability of the halting problem, so the fact that Theorem~\ref{KcompOPT} 
shows that the sets ${\rm OPT}_A$ are oracle-computable given an oracle that
can compute Kolmogorov complexity (hence that can solve the halting problem)
shows that if the sets ${\rm OPT}_A$ are actually non-computable, the source of
the non-computability would indeed be of this nature. 

\smallskip

It is clear that in all cases, the fact that existing tests of optimality (such as Cohn-Elkies bounds)
are sufficient but not necessary makes it impossible to derive from them either a direct
proof of computability or a proof of non-computability that shows the necessity of
the halting problem, or of another non-decidable problem.

\end{document}